\documentclass[aps,prb,twocolumn,superscriptaddress]{revtex4}
\usepackage{graphicx}

\begin{document}

\title{
Possible realization of an ideal quantum computer in Josephson junction array
} 
% with topologically protected degenerate ground states. }
\author{L.B. Ioffe} 

\altaffiliation{Landau Institute for Theoretical Physics, Kosygina 2, Moscow, 117940 Russia}

\affiliation{Center for Materials Theory,  
Department of Physics and Astronomy, Rutgers University 
136 Frelinghuysen Rd, Piscataway NJ 08854 USA}

\author{M.V. Feigel'man}
\affiliation{Landau Institute for Theoretical Physics, Kosygina 2, Moscow, 117940 Russia}

\begin{abstract}
We introduce a new class of Josephson arrays which have non-trivial topology and 
exhibit a novel state at low temperatures. This state is characterized by 
long range order in a \textit{two} Cooper pair condensate and by a discrete topological 
order parameter. These arrays have degenerate ground states with
this degeneracy 'protected' from the external perturbations (and noise) by the topological 
order parameter. We show that in ideal conditions the low order effect of the
external perturbations on this degeneracy is exactly zero and that deviations from ideality 
lead to only exponentially small effects of perturbations. We argue that this system provides 
a physical implementation of an ideal quantum computer with a built in error correction and 
show that even a small array exhibits interesting physical properties such as superconductivity
with double charge, $4e$, and extremely long decoherence times.  
\end{abstract}

\maketitle

\section{Introduction}

Quantum computing \cite{Steane1998,Ekert1996} is in principle  a very powerful technique 
for solving classic 'hard' problems such as factorizing large numbers 
\cite{Shor1994} or sorting large lists \cite{Grover1996}. The remarkable discovery of quantum 
error correction algorithms \cite{Shor1995} shows there is no problem of 
principle involved in building a functioning quantum computer. However, implementation still 
seems dauntingly difficult: the essential ingredient of a quantum computer is a quantum system 
with $2^K$ (with $K \sim 104-106$) quantum states which are degenerate (or nearly so) in the 
absence of external perturbations and are insensitive to the 'random' fluctuations which exist 
in every real system, but which may be manipulated by controlled external fields with errors 
less $10^{-4}-10^{-6}$ (big system sizes, $K$, are needed to correct the errors, for smaller errors 
the size of the system, $K$, gets much smaller) \cite{Preskill1998}. 
Insensitivity to random fluctuations means that any coupling to the external
environment neither induces transitions among these  $2^K$ states nor changes the phase of one 
state with respect another. Mathematically, this means that one requires a system whose 
Hilbert space contains a $2^K$-dimensional subspace (called 'the protected subspace'\cite{Kitaev1997})
within which any local operator $\hat{O}$ has (to a high accuracy) only
state-independent diagonal matrix elements:
$\langle n | \hat{O} | m \rangle = O_0 \delta_{mn} + o(\exp(-L))$ where $L$ is a parameter
such as the system size that can be made as large as desired. It has been
very difficult to design a system which meets these criteria.
Many physical systems (for example, spin glasses \cite{SG}) exhibit exponentially many essentially 
degenerate states, not connected to each other by local operators. However, the requirement 
that all diagonal matrix elements are equal (up to vanishingly small terms) is highly
nontrivial and puts such systems in a completely new class. Parenthetically we note that such 
systems were discussed in philosophical terms by I. Kant (who termed them noumenons, in his 
thinking the noumenal world is impenetrable but contains comprehensible information) 
in \cite{Kant1781}.

One very attractive possibility, proposed in an important paper by Kitaev \cite{Kitaev1997} 
and developed further in \cite{Kitaev_etal} involves a protected subspace 
\cite{Wen1990,Wen1991} created by a topological 
degeneracy of the ground state. Typically such degeneracy happens if the system has a conservation
law such as the conservation of the \textit{parity} of the number of 'particles' along some long 
contour. Physically, it is clear that two states that differ only by the parity of some big number that 
can not be obtained from any local measurement are very similar to each other.
The model proposed in \cite{Kitaev1997} has been shown to exhibit many properties of
the ideal quantum computer; however before now no robust and practical implementation was known. 
In a recent paper we and others proposed a Josephson junction network which is an implementation
of a similar model with protected degeneracy and which is possible (although difficult) to build 
in the laboratory \cite{Ioffe2002}. 

In the present paper we develop ideas of~\cite{Ioffe2002} proposing a new Josephson
junction network that has a number of practical advantages. 
(i) This network operates in a phase regime (i.e. when Josephson energy is larger 
than the charging energy), which reduces undesired effects of parasitic stray charges.
(ii) All Josephson junctions in this array are similar which should simplify the 
fabrication process.  
(iii) This system has $2^K$ degenerate ground states 'protected' to even higher extent than 
in~\cite{Ioffe2002}: matrix elements of local operators scale as $\varepsilon^L$, 
where $\varepsilon \leq 0.1$ is a measure of non-ideality of the system's 
fabrication (e.g. the spread of critical currents of different Josephson junctions and
geomertical areas of different elementary cells in the network) 
(iv) The new array does not require a fine tuning of its parameters into a narrow region.
The relevant degrees of freedom of this new array are described by the model analogous
to the one proposed in \cite{Kitaev1997}.  Even when such system is small and contains
only a few 'protected' states its physical properties are remarkable: it is a superconductor
with the elementary charge $4e$ and the decoherence time of the protected states can be made 
macroscopic allowing 'echo' experiments. 

Below we first describe the physical array, identify its relevant low energy degrees of freedom
and the mathematical model which describes their dynamics. We then show how the protected states
appear in this model, derive the parameters of the model and identify various corrections
appearing in a real physical system and their effects. Finally, we discuss how one can manipulate
these states in a putative quantum computer and the physical properties expected in a small
arrays of this type.  

\section{Array}

The basic building block of the lattice is a rhombus made of four Josephson
junctions with each side of the rhombus containing one Josephson contact, these rhombi 
form a hexagonal lattice as shown in Fig.~1.  We denote
the centers of the hexagons by letters $a,b\ldots$ and the individual
rhombi by $(ab),(cd) \ldots$, because each rhombus is one-to-one correspondence with the 
link $(ab)$ between the sites of the triangular lattice dual to the hexagonal lattice.  
The lattice is placed in a uniform 
magnetic field so that the flux through each rhombus is $\Phi_0/2$. The geometry
is chosen in such a way that the flux, $\Phi_s$ through each David's star is a 
half-integer multiple of $\Phi_0$: $\Phi = (n_s+\frac{1}{2}) \Phi_0$. 
\footnote{
The flux $\Phi_s$ can be also chosen so that it is an integer multiple of $\Phi_0$: this
would not change significantly the final results but would change intermediate 
arguments and make them longer, so for clarity we discuss in detail only the 
half-integer case here.
Note, however, that the main quantitatve effect of this alternative choice of the flux is
beneficial: it would push up the phase transition line separating the topological and 
superconducting phases shown in Fig.~3 for half integer case.
}
Finally, globally the lattice contains a number, $K$, of big openings (the size of 
the opening is much larger than the lattice constant, a lattice with $K=1$ is shown in Fig.~1a). 
The dimension of the protected space will be shown to be equal $2^K$. The system is characterized
by the Josephson energy, $E_J=\frac{\hbar}{2e}I_c$, of each contact and by the capacitance 
matrix of the islands (vertices of the lattice). We shall assume (as is usually the case) that the 
capacitance matrix is dominated by the capacitances of individual junctions, we write the charging
energy as $E_C=\frac{e2}{2C}$. The 'phase regime' of the network mentioned above
implies that $E_J > E_C$. The whole system is described by the Lagrangian
\begin{equation}
\mathcal{L} =   \sum_{(ij)} \frac{1}{16 E_C} (\dot{\phi_i} - \dot{\phi_j})2 + 
	E_J \cos(\phi_i-\phi_j-a_{ij})  
\label{L}
\end{equation}
where $\phi_i$ are the phases of individual islands
and $a_{ij}$ are chosen to produce the correct magnetic fluxes.
The Lagrangian (\ref{L}) contains only gauge invariant phase differences, 
$\phi_{ij}=\phi_i-\phi_j-a_{ij}$,
so it will be convenient sometimes to treat them as independent variables satisfying the 
constraint $\sum_{\Gamma} \phi_{ij} = 2\pi \Phi_{\Gamma} / \Phi_0 + 2\pi n$ 
where the sum is taken over closed loop $\Gamma$ and $n$ is arbitrary integer.

\begin{figure}[ht]
\includegraphics[width=3.2in]{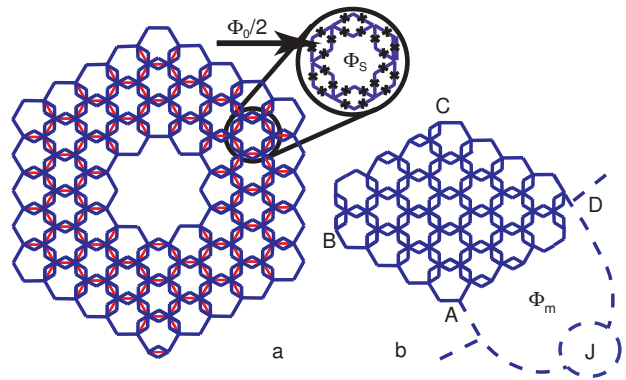}
\caption{
Examples of the proposed Josephson junction array. 
Thick lines show superconductive wires, each wire contains one Josephson junction
as shown in detailed view of one hexagon. 
The array is put in magnetic field such that the flux through each elementary rhombus and 
through each David's star (inscribed in each hexagon) is half integer. 
Thin lines show the effective bonds formed by the elementary rhombi. The Josephson
coupling provided by these bonds is $\pi$-periodic. 
a. Array with one opening, generally the effective number of qubits, $K$ is equal to the 
number of openings. The choice of boundary condition shown here makes superconducting phase
unique along the entire length of the outer (inner) boundary, the state of the entire
boundary is described by a single degree of freedom. The topological order parameter controls 
the phase difference between inner and outer boundaries. Each boundary includes one rhombus to
allow experiments with flux penetration; magnetic flux through the opening
is assumed to be $\frac{\Phi_0}{2}(\frac{1}{2} +m)$ with any integer $m$.
b. With this choice of boundary circuits the phase is unique only inside the sectors $AB$ and 
$CD$ of the boundary; the topological degree of freedom controls the difference between the
phases of these boundaries. This allows a simpler setup of the experimental test for the signatures
of the ground state described in the text, e.g. by a SQUID interference experiment 
sketched here that involves a measuring loop with flux $\Phi_m$ and a very weak 
junction $J$ balancing the array.   
}
\end{figure}

	As will become clear below, it is crucial that the degrees of freedom at
the boundary have dynamics identical to those in the bulk. To ensure this one needs to add 
additional superconducting wires and Josephson junctions at the boundary. There are
a few ways to do this, two examples are shown in Fig.~1a and Fig.~1b: type I boundary 
(entire length of boundaries in Fig.~1a, parts $AB$, $CD$) and type II boundary ($BC$,
$AD$).  
For both types of boundaries one needs to include in each boundary loop the flux which 
is equal to $Z_b*\pi/2$ where $Z_b$ is coordination number of the dual triangular lattice site. 
For instance, for the four coordinated boundary sites one needs to enclose the integer flux 
in these contours. 
In type I boundary the entire boundary corresponds to one degree of freedom  (phase
at some point)
while type II boundary includes many rhombi so it contains many degrees of freedom.

Note that each (inner and outer) boundary shown in Fig.~1a contains one rhombus; 
we included it to allow flux to enter and exit through the boundary when it is energetically
favorable.

\begin{figure}[ht]
\includegraphics[width=3.2in]{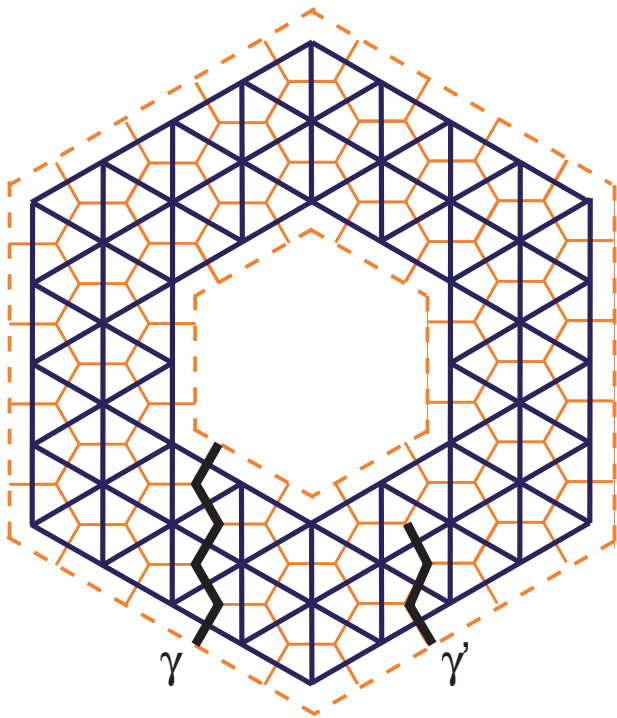}
\caption{
The spin degrees of freedom describing the state of the elementary rhombi are
located on the bonds of the triangular lattice (shown in thick lines)
while the constraints are defined on the sites of this lattice. The dashed line
indicates the boundary condition imposed by a physical circuitry shown in Fig.~1a.
Contours $\gamma$ and $\gamma'$ are used in the construction of topological order
parameter and excitations. 
}
\end{figure}

\section{Ground state, excitations and topological order}
In order to identify the relevant degrees of freedom in this highly frustrated system we 
consider first an individual rhombus. As a function of the gauge invariant phase difference 
between the far ends of the rhombus the potential energy is 
\begin{equation}
U(\phi_{ij})= - 2 E_J ( |\cos(\phi_{ij}/2)| + |\sin(\phi_{ij}/2)| ).
\label{U}
\end{equation} 
This energy has two equivalent minima, at $\phi_{ij} = \pm \pi/2$ which can be used
to construct elementary unprotected qubit, see \cite{Ioffe2001}. In each of these states
the phase changes by $\pm \pi/4$ in each junction clockwise around the rhombus. We denote these
states as $| \uparrow \rangle$ and $| \downarrow \rangle$ respectively. 
In the limit of large Josephson energy the space of low energy states of the full lattice is
described by these binary degrees of freedom, the set of operators acting on these states 
is given by Pauli matrices $\sigma^{x,y,z}_{ab}$. 
We now combine these rhombi into hexagons forming the lattice shown in Fig.~1.
This gives another condition: the sum of phase differences around the hexagon should
equal to the flux, $\Phi_s$ through each David's star inscribed in this hexagon. 
The choice $\phi_{ij} = \pi/2$ is consistent with flux $\Phi_s$ that is equal to a 
half integer number of flux quanta.  
This state minimizes the potential energy (\ref{U}) of the system. 
This is, however, not the only choice. Although flipping 
the phase of one dimer changes the phase flux around the star by $\pi$ and thus is 
prohibited, flipping two, four and six rhombi is allowed; generally 
the low energy configurations of $U(\phi)$ satisfy the constraint 
\begin{equation}
\hat{P}_a = \prod_b \sigma^z_{ab} = 1
\label{P}
\end{equation} 
where the product runs over all neighbors, $b$, of site $a$.
The number of (classical) states satisfying the constraint (\ref{P}) is still huge:
the corresponding configurational entropy is extensive (proportional to the number of
sites). 
We now consider the charging energy of the contacts, which results in the quantum dynamics 
of the system. We show that it reduces this degeneracy to a much smaller number $2^K$. 
The dynamics of the individual rhombus is described by a simple Hamiltonian 
$H=\tilde{t} \sigma_x$ but the dynamics of a rhombus 
embedded in the array is different because individual flips are not compatible
with the constraint (\ref{P}). The simplest dynamics compatible with (\ref{P}) contains
flips of three rhombi belonging to the elementary triangle, $(a,b,c)$, 
$\hat{Q}_{(abc)} = \sigma^x_{ab} \sigma^x_{bc} \sigma^x_{ca}$ and therefore the
simplest quantum Hamiltonian operating on the subspace defined by (\ref{P}) is
\begin{equation}
H  = - r \sum_{(abc)} Q_{(abc)}
\label{H_model}
\end{equation}
We discuss the derivation of the coefficient $r$ in this Hamiltonian and the correction 
terms and their effects below but first we solve the simplified model 
(\ref{P},\ref{H_model}) and show that its ground state is ``protected'' in the sense 
described above and that excitations are separated by the gap.
\footnote{In a rotated basis $\sigma^x\rightarrow \sigma^z$, $\sigma^z\rightarrow \sigma^x$ this 
model is reduced to a special case of $Z_2$ lattice gauge theory \cite{Wegner1971,Balian1975}
which contains only magnetic term in the Hamiltonian
with the constraint (\ref{P}) playing a role a gauge invariance condition.
} 

Clearly, it is very important that the constraint is imposed on all sites,
including boundaries. Evidently, some boundary hexagons are only partially complete
but the constraint should be still imposed on the corresponding sites of the corresponding 
triangular lattice. This is ensured by additional superconducting wires that close the
boundary hexagons in Fig.~1.  
 
We note that constraint operators commute not only with the full Hamiltonian but also 
with individual $\hat{Q}_{(abc)}$: $[\hat{P}_a, \hat{Q}_{(abc)}]=0$.
The Hamiltonian (\ref{H_model}) without constraint has an obvious ground state, 
$|0\rangle$, in which $\sigma^x_{ab}=1$ for all rhombi. This ground state, however, 
violates the constraint. This can be fixed noting that since operators $\hat{P}_a$ 
commute with the Hamiltonian, any state obtained from $|0\rangle$ by acting on it 
by $\hat{P}_a$ is also a ground state. We can now construct a true ground state 
satisfying the constraint by
\begin{equation}
|G\rangle = \prod_a \frac{1+\hat{P}_a}{\sqrt{2}} |0 \rangle
\label{G}
\end{equation}
Here $\frac{1+\hat{P}_a}{\sqrt{2}}$ is a projector onto the subspace satisfying the 
constraint at site $a$ and preserving the normalization.  

Obviously, the Hamiltonian (\ref{H_model}) commutes with any product of $\hat{P}_a$ 
which is equal to to the product of $\sigma^z_{ab}$ operators around a set of
closed loops. These integrals of motion are fixed by the constraint.  However, 
for a topologically non-trivial system there appear a number of other integrals
of motion. For a system with $K$ openings a product of $\sigma^z_{ab}$ operators 
along contour, $\gamma$ that begins at one opening and ends at another (or at the outer
boundary, see Fig.~2)
\begin{equation}
\hat{T_q} = \prod_{(\gamma_q)} \sigma^z_{ab}
\label{T}
\end{equation}
commutes with Hamiltonian and is not fixed by the constraint. 
Physically these operators count the parity of 'up' rhombi along such contour.
The presence of these operators results in the degeneracy of the ground state. 
Note that multiplying such operator by an appropriate $\hat{P}_a$ gives similar operator
defined on the shifted contour so all topologically equivalent contours give 
one new integral of motion. Further, multiplying two operators defined along the contours
beginning at the same (e.g. outer) boundary and ending in different openings, $A$, $B$ is 
equivalent to the operator defined on the contour leading from $A$ to $B$, so 
the independent operators can be defined (e.g.) by the set of contours that begin at one 
opening and ends at the outer boundary. 
The state $|G\rangle$ constructed above is not an eigenstate of these operators but this
can be fixed defining 
\begin{equation}
|G_{f}\rangle = \prod_q \frac{1 + c_q \hat{T_q}}{\sqrt{2}} |G\rangle
\label{G_f}
\end{equation}
where $c_q = \pm 1$ is the eigenvalue of $\hat{T_q}$ operator defined on contour $\gamma_q$. 
Equation (\ref{G_f}) is the final expression for the ground state eigenfunctions. 

Construction of the excitations is similar to the construction of the ground state. 
First, one notices that since all operators $\hat{Q}_{abc}$ commute with each other and
with the constraints, any state of the system can be characterized by the eigenvalues 
($Q_{abc}= \pm 1$) of $\hat{Q}_{abc}$. The lowest excited state correspond to only one $Q_{abc}$
being $-1$. Notice that a simple flip of one rhombus (by operator $\sigma^z_{(ab)}$ 
somewhere in the system changes the sign of \textit{two} $Q_{abc}$ eigenvalues 
corresponding to two triangles to which it belongs. To change only one  $Q_{abc}$ one needs to
consider a continuous string of these flip operators starting from the boundary: 
$| (abc) \rangle = v_{(abc)} |0\rangle$ with $ v_{(abc)} = \prod_{\gamma'} \sigma^z_{(cd)}$ where 
the product is over all rhombi $(cd)$ that belong to the path, $\gamma'$, that begins at the boundary 
and ends at $(abc)$ (see Fig.~2 which shows one such path). 
This operator changes the sign of only one ${Q}_{abc}$, the one that corresponds to 
the 'last' triangle. This construction does not satisfy the constraint, so we have to 
apply the same 'fix' as for the ground state construction above 
\begin{equation}
|v_{(abc)}\rangle = \prod_q \frac{1+c_q \hat{T_q}}{\sqrt{2}} 
\prod_a \frac{1+\hat{P}_a}{\sqrt{2}} v_{(abc)} |0 \rangle
\label{v_abc}
\end{equation} 
to get the final expression for the lowest energy excitations. The energy of each excitation
is $2r$. Note that a single 
flip excitation at a rhombus $(ab)$ can be viewed as a combination of two elementary excitations 
located at the centers of the triangles to which rhombus $(ab)$ belongs and has twice their energy. 
Generally, all excited states of the model (\ref{H_model}) can be characterized as a number
of elementary excitations (\ref{v_abc}), so they give exact quasiparticle basis.
Note that creation of a quasiparticle at one boundary and moving it to another is equivalent to
the $\hat{T}_q$ operator, so this process acts as $\tau^z_q$ in the space of the 
$2^K$ degenerate ground states. As will be shown below, in the physical system of
Josephson junctions these excitations carry charge $2e$ so that $\tau^z_q$ process 
is equivalent to the charge $2e$ transfer from one boundary to another. 

Consider now the matrix elements, $O_{\alpha \beta}=\langle G_\alpha | \hat{O} | G_\beta 
\rangle$ of a local operator, $\hat{O}$, between two ground states, e.g.
of an operator that is composed of a small number of $\sigma_{ab}$.  
To evaluate this matrix element we first project a general operator onto the
space that satisfies the constraint: 
$\hat{O} \rightarrow \mathcal{P} \hat{O} \mathcal{P} $ where 
$\mathcal{P}=\prod_a \frac{1+\hat{P}_a}{2}$. The new (projected) operator is also local, it has
the same matrix elements between ground states but it commutes with all $\hat{P}_a$. Since it is
local it can be represented as a product of $\sigma_z$ and $\hat{Q}$ operators which implies that 
it also commutes with all $\hat{T_q}$. Thus, its matrix elements between different
states are exactly zero. Further, using the fact that it commutes with $\hat{P}_a$ and $\hat{T_q}$
we write the difference between its diagonal elements evaluated between the states that differ
by a parity over contour $q$ as 
 \begin{equation}
O_{+}-O_{-} = \langle 0 | \prod_i \frac{1+\hat{P}_i}{\sqrt{2}} \hat{T_q} \hat{O} | 0 \rangle 
\label{O_pm}
\end{equation}
This equation can be viewed as a sum of products of $\sigma_z$ operators. Clearly to 
get a non-zero contribution each $\sigma^z$ should enter even number of times. 
Each $\hat{P}$ contains a closed loop of six $\sigma^z$ operators, so any product of
these terms is also a collection of a closed loops of $\sigma^z$. In contrast to it, 
operator $\hat{T}_q$  contains a product of $\sigma^z$ operators along the loop $\gamma$, 
so the product of them contains a string of $\sigma_z$ operators along the contour 
that is topologically equivalent to $\gamma$. 
Thus, one gets a non-zero $O_{+}-O_{-}$ only for the operators $\hat{O}$ which 
contain a string of $\sigma^z$ operators along the loop that is topologically 
equivalent to $\gamma$ which is impossible for a local operator.
Thus, we conclude that for this model all non-diagonal matrix elements of a local operator are
\textbf{exactly} zero while all diagonal are \textbf{exactly} equal. 

\section{Effect of physical perturbations}

We now come back to the original physical system described by the Lagrangian
(\ref{L}) and derive the parameters of the model (\ref{H_model}) and discuss
the most important corrections to it and their effect. We begin with the
derivation. In the limit of small charging energy the flip of three rhombi 
occurs by a virtual process in which the phase, $\phi_i$ at one (6-coordinated) 
island, $i$, changes by $\pi$.
In the quasiclassical limit the phase differences on the individual junctions
are $\phi_{ind}=\pm \pi/4$;
the leading quantum process changes the phase on one junction by $3\pi/2$ and
on others by $-\pi/2$ changing the phase across the rhombus $\phi \rightarrow \phi+\pi$. 
%and the effective Lagrangian of each rhombus can be written in terms of only the phase
%$\phi$:
%\begin{equation}
%L_{rhombus} = \frac{1}{16 E_C} (\dot{\phi})2 + U(\phi)
%\label{L_rhombus}
%\end{equation} 
The phase differences, $\phi$, satisfy the constraint that the sum of them over the
closed loops remain $2\pi (n+\Phi_s/\Phi_0)$. The simplest such process preserves 
the symmetry of the lattice, and changes simultaneously the phase differences on 
the three rhombi containing island $i$  keeping all other phases constant. 
The action for such process is three times the action of 
elementary transitions of individual rhombi, $S_0$:
\begin{equation}
r \approx E_J^{3/4} E_C^{1/4} \exp(-3 S_0), \; \; S_0=1.61 \sqrt{E_J/E_C}
\label{r}
\end{equation}
In the alternative process the phase differences between $i$ and other islands
change in turn, via high energy intermediate state in which one phase difference
has changed while others remained close to their original values. 
The estimate for this action shows that it is larger than $3S_0$, so (\ref{r}) gives
the dominating contribution. 
There are in fact many processes that contribute to this transition:
the phase of island $i$ can change by $\pm \pi$ and in addition
in each rombus one can choose arbitrary the junction in which the phase
changes by $\pm 3\pi/2$; the amplitude of all these processes should be added. 
This does not change the result qualitatively unless these amplitudes exactly 
cancel each other, which happens only if the charge of the island is exactly half-integer 
(because phase and charge are conjugate the amplitude difference of the processes that 
are different by $2\pi$ is $\exp(2\pi iq)$ ). 
We assume that in a generic case this cancellation does not occur. External 
electrical fields (created by e.g. stray charges) might induce non-integer 
charges on each island which would lead to a randomness in the phase and
amplitude of $r$. The phase of $r$ can be eliminated by a proper gauge transformation 
$|\uparrow\rangle_{ab} \rightarrow e^{i\alpha_{ab}}|\uparrow\rangle_{ab}$
and has no effect at all. The amplitude variations result in a position
dependent quasiparticle energy. 

We now consider the corrections to the model (\ref{H_model}). 
One important source of corrections is the difference of the actual magnetic flux
through each rhombus from the ideal value $\Phi_0/2$. If this difference is small
it leads to the bias of 'up' versus 'down' states, their energy difference becomes
$2 \epsilon = 2\pi \sqrt{2} \frac{\delta \Phi_d}{\Phi_0} E_J$. Similarly, the difference
of the actual flux through David's star and the difference in the Josephson energies
of individual contacts leads to the interaction between 'up' states:
\begin{equation}
\delta H_1 =  \sum_{(ab)} V_{ab} \sigma^z_{ab} +  
	\sum_{(ab),(cd)} V_{(ab),(cd)} \sigma^z_{ab} \sigma^z_{cd}
\label{deltaH_1}
\end{equation}
where $V_{ab} = \epsilon$ for uniform field deviating slightly from the ideal value
and $V_{(ab),(cd)} \neq 0$ for rhombi belonging to the same hexagon. 
Consider now the effect of perturbations described by $\delta H_1$, Eq. (\ref{deltaH_1}). These
terms commute with the constraint but do not commute with the main term, $H$, so the ground state 
is no longer $|G_{\pm} \rangle$. In other words, these terms create excitations (\ref{v_abc}) 
and give them kinetic energy. In the leading order of the perturbation theory the ground state 
becomes $| G_{\pm}\rangle + \frac{\epsilon}{4r} \sum_{(ab)} \sigma^z_{(ab)} |G_{i\pm}\rangle$
Qualitatively, it corresponds to the appearance of virtual pairs of quasiparticles in the ground 
state. The density of these quasiparticles is $\frac{\epsilon}{r}$. As long as these quasiparticles 
do not form a topologically non-trivial string all previous conclusions remain valid. However, 
there is a non-zero amplitude to form such string  - it is now exponential in 
the system size. With exponential accuracy this amplitude is $(\frac{\epsilon}{2r})^L$ which 
leads to an energy splitting of the two ground state levels and the matrix elements of
typical local operators of the same order 
$E_{+} - E_{-} \sim O_{+}-O_{-} \sim (\frac{\epsilon}{2r})^L$.

The physical meaning of the $v_{(abc)}$ excitations become more clear if one consider the effect of 
the addition of one $\sigma^z$ operator to the end of the string defining the quasiparticle:
it results in the charge transfer of $2e$ across this last rhombus. To prove this, note that the 
wave function of a superconductor corresponding to the state which is symmetric combination of 
$|\uparrow \rangle$ and $|\downarrow \rangle$  is periodic with period $\pi$ and thus corresponds 
to charge which is multiple of $4e$ while the antisymmetric corresponds to charge $(2n+ 1)2e$. 
The action of one $\sigma_z$ induces the transition between these states and thus transfers the 
charge $2e$. Thus, these excitations carry charge $2e$. Note that continuous degrees of freedom 
are characterized by the long range order in $\cos(2\phi)$ and thus correspond to the condensation 
of pairs of Cooper pairs. In other words, this system superconducts with elementary charge $4e$ 
and has a gap, $2r$, to the excitations carrying charge $2e$. Similar pairing of Cooper pairs was shown 
to occur in a chain of rhombi in a recent paper \cite{Doucot2002}; formation of a classical 
superconductive state with effective charge $6e$ in frustrated Kagome wire network was predicted 
in \cite{Huse2001}.

The model (\ref{H_model}) completely ignores the processes that violate the constraint at each 
hexagon. Such processes might violate the conservation of the topological invariants $\hat{T}_q$
and thus are important for a long time dynamics of the ground state manifold. In order to consider
these processes we need to go back to the full description involving the continuous superconducting
phases $\phi_i$. Since potential energy (\ref{U}) is periodic in $\pi$ it is convenient to separate 
the degrees of freedom into continuous part (defined modulus $\pi$) and discrete parts. Continuous 
parts have a long range order: $<\cos(2\phi_0 - 2\phi_r)> \sim 1$. The elementary excitations of the
continuous degrees of freedom are harmonic oscillations and vortices. The harmonic oscillations 
interact with discrete degrees of freedom only through the local currents that they generate, 
further the potential (\ref{U}) is very close to the quadratic, so we conclude that they are 
practically decoupled from the rest of the system. In contrast to this vortices have an important 
effect. By construction,
the elementary vortex carries flux $\pi$ in this problem. Consider the structure of these vortices
in a greater detail. The superconducting phase should change by $0$ or $2\pi$ when one moves
around a closed loop. In a half vortex this is achieved if the gradual change by $\pi$ is 
compensated (or augmented) by a discrete change by $\pi$ on a string of rhombi which costs no
energy. Thus, from the viewpoint of discrete degrees of freedom the position of the vortex is the
hexagon where constraint (\ref{P}) is violated.   
The energy of the vortex is found from the usual arguments
\begin{equation}
E_v(R) = \frac{\pi E_J}{4 \sqrt{6} } 
\left( \ln\left(R \right) + c \right), \; \; c\approx 1.2; 
\label{E_vortex}
\end{equation}         
it is logarithmic in the vortex size, $R$. The process that changes the topological invariant, 
$\hat{T}_q$
is the one in which one half vortex completes a circle around an opening. The amplitude of 
such process is exponentially small: $(\tilde{t}/E_v(D))^\Lambda$ where $\tilde{t}$ is 
the amplitude to flip one rhombus and $\Lambda$ is the length of the shortest path around the opening. 
In the quasiclassical limit the amplitude $\tilde{t}$ can be estimated analogously to (\ref{r}): 
$\tilde{t} \sim \sqrt{E_J E_Q} \exp(-S_0)$. 
The half vortices would appear in a realistic system if the flux through each hexagon is 
systematically different from the ideal half integer value. The presence of free vortices 
destroys topological invariants, so a realistic system should either be not too large (
so that deviations of the total flux do not induce free vortices) or these vortices should be 
localized in prepared traps (e.g. David's stars with fluxes slightly larger or smaller than $\Phi_s$). 
If the absence of half vortices the model is equivalent to the Kitaev model 
\cite{Kitaev1997} placed on triangular lattice in the limit of the infinite energy of the 
excitation violating the constraints. 

%These excitations are different: in Josephson junction
%arrays they are vortices and thus have logarithmic energy.  

Quantitatively, the expression for the parameters of the model (\ref{H_model}) become exact only if
$E_J \gg E_C$. One expects, however, that the qualitative conclusions remain the same and the 
formulas derived above provide good estimates of the scales even for $E_J \sim E_C$, provided that 
charging energy is not so large as to result in a phase transition to a different phase. 
One expects this transition to occur at $E_c^* = \eta E_J$ with $\eta \sim 1$ which exact value can be
reliably determined only from numerical simulations.  Practically, since the perturbations induced 
by flux deviations from $\Phi_0$ are proportional to $\frac{\delta \Phi}{\Phi_0} \frac{E_J}{r}$ 
and $r$ becomes exponentially small at small $E_C$, the optimal choice of the parameters for the 
physical system is $E_C \approx E_C^*$. We show the schematics of the phase diagram in Fig.~3.
\footnote{
We assume that transition to insulating phase is direct,
another alternative is the intermediate phase in which the energy of the
vortex becomes finite instead of being logarithmic. If this phase indeed exists it 
is likely to have properties more similar to the one discussed in \cite{Kitaev1997}. 
}
The 'topological' phase is stable in a significant part of the phase diagram, further since
the vortex excitations have logarithmic energy, we expect that this phase survives at
finite temperatures as well. In the thermodynamic limit, at $T\neq 0$ one gets a finite density 
of $2e$ carrying excitations ($n_v \sim \exp(-2r/T)$) but the 
vortices remain absent as long as temperature is below BKT-like 
depairing transition for half-vortices.

\begin{figure}[ht]
\includegraphics[width=2in]{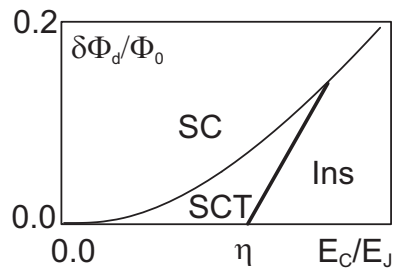}
\caption{
Schematic of the phase diagram for half integer $\Phi_s$ at low temperatures: $\delta \Phi_d$ 
is the deviation of the magnetic flux through each rhombus from its ideal value. SC stands for 
usual superconducting phase,
SCT for the phase with $\cos(2\phi)$ long range order of the continuous degrees of freedom 
and discrete topological order parameter discussed extensively in the bulk of the paper. 
The SCT phase and SC phase are separated by 2D quantum Ising phase transition. 
}
\end{figure}

\section{Quantum manipulations}

We now discuss the manipulation of the protected states formed in this system. First, we note that
the topological invariant $\hat{T_q}$ has a simple physical meaning - it is the total phase 
difference (modulus $2 \pi$) between the inner and outer boundaries. This means that measuring this
phase difference measures the state of the qubit. Also, introducing a weak coupling between these
boundaries by a very weak Josephson circuit (characterized by a small energy $\epsilon_J$) would 
change the phase of these states in a controllable manner, e.g. in a unitary transformation 
$U^z=\exp(i \epsilon_J t \tau^z_q)$. 
The transformation coupling two qubits can be obtained if one introduces a weak Josephson circuit
that connects two different inner boundaries (corresponding to different qubits).
Analogously, the virtual process involving half vortex motion around the opening gives
the tunneling amplitude, $\epsilon_t$ between topological sectors, e.g. unitary 
transformation $U^x=\exp(i t \epsilon_t \tau^x_q)$. This tunneling can be controlled by magnetic
field if the system is prepared with some number of vortices that are pinned in the idle state
in a special plaquette where the flux is integer. The slow (adiabatic) change of this flux to
towards a normal (half-integer) value would release the vortex and result in the transitions
between topological sectors with $\epsilon_t \sim \tilde{t}/D2$  

These operations are analogous to usual operations on a qubit and are prone 
to usual source of errors. This system, however, allows another type of operation that are 
naturally discrete. As we show above the transmission of the elementary quasiparticle across the 
system changes its state by $\tau^z_q$.  This implies that a discrete process of one pair 
transfer across the system is equivalent to the $\tau^z_q$ transformation. Similarly, a controlled
process in which a vortex is moved around a hole results in a discrete $\tau^x_q$ transformation.
Moreover, this system allow one to make discrete transformations such as $\sqrt{\tau^{x,z}}$. 
Consider, for instance, a process in which by changing the total magnetic flux through the system
one half vortex is placed in a center of the system shown in Fig.~1b and then released. It can 
escape through the left or through the right boundary, in one case the state does not change, 
in another it changes by $\tau_x$. The amplitudes add resulting in the operation 
$\frac{1+i\tau^x}{\sqrt{2}}$. 
Analogously, using the electrostatic gate(s) to pump one charge $2e$ from one boundary to the
island in the center of the system and then releasing it results in a $\frac{1+i\tau^z}{\sqrt{2}}$ 
transformation. This type of processes allow a straightforward generalization for the array
with many holes: there extra half vortex or charge should be placed at equal distances from the
inner and outer boundaries. 

\section{Physical properties of small arrays}

Even without these applications for quantum computation the physical properties of this array 
are remarkable: it exhibits a long range order in the square of the usual superconducting order
parameter: $\langle \cos(2 (\phi_0-\phi_r)) \rangle \sim 1$ without the usual order:
$\langle \cos(\phi_0-\phi_r) \rangle = 0$; the charge transferred through the system is quantized
in the units of $4e$. This can be tested in a interference experiment sketched in Fig.~1b,
as a function of external flux, $\Phi_m$ the supercurrent through the loop should be periodic with 
half the usual period. This simpler array can be also used for 
a kind of 'spin-echo' experiment:
applying two consecutive operations $\frac{1+i\tau_x}{\sqrt{2}}$ described above should give
again a unique classical state while applying only one of them should result in a quantum
superposition of two states with equal weight. 

The echo experiment can be used to measure the decoherence time in this system. 
Generally one distinguishes processes that flip the classical states and the ones
that change their relative phases. In NMR literature the former are referred to
as transverse relaxation and the latter as longitudinal one. The transverse
relaxation occurs when a vortex is created and then moved around the opening by an
external noise. Assuming a thermal noise, we estimate the rate of this process 
$\tau_\perp^{-1}\sim \tilde{t} \exp(-E_V(L)/T)$. 
Similarly, the transfer of a quasiparticle from the outer to the inner boundary
changes the relative phase of the two states, leading to a longitudinal relaxation.
This rate is proportinal to the density of quasiparticles, 
$\tau_\parallel^{-1} = {\cal R} \exp(-2r/T)$. The coefficient ${\cal{R}}$
depends on the details of the physical system. 
In an ideal system with some nonzero uniform value of $\epsilon$
(defined above  (\ref{deltaH_1})) quasiparticles are delocalized and 
${\cal R} \sim \epsilon/L2$. Random deviations of fluxes $\Phi_r$ from half-integer value 
produce randomness in $\epsilon$, in which case one expects Anderson
localization of quasiparticles due to {\it off-diagonal} disorder,
with localization length  of the order of lattice spacing, thus
 ${\cal R} \sim \bar{\epsilon}\exp(-c L)$  with $c \sim 1$,
and $\bar{\epsilon}$ is the typical value of $\epsilon$.
Stray charges induce randomness in the values of $r$, i.e. add 
some {\it diagonal}  disorder. When the random part of $r$, $\delta r$ becomes
larger than $\bar{\epsilon}$ the localization becomes stronger:
${\cal R} \sim \bar{\epsilon}(\bar{\epsilon}/\bar{\delta r})^L$
where $\bar{\delta r}$ is the typical value of $\delta r$. 
Upon a further increase of stray charge field there appear rare sites where $r_i$ is
much smaller than an average value. Such site acts as an additional openings in 
the system. If the density of these sites is significant, the effective length
that controls the decoherence becomes the distance between these sites. 
For typical $E_V(L)\approx E_J\approx 2K$ the transverse relaxation 
time reaches seconds for $T \sim 0.1 \;K$ while realistic 
$\epsilon/r \sim 0.1$ imply that due to a quasiparticle localization in a random case 
the longitudinal relaxation reaches the same scale for systems of size $L \sim 10$; 
note that temperature $T$ has to be only somewhat lower than the excitation gap, $2r$, 
in order to make the longitudinal rate low.

Most properties of the array are only weakly sensitive to the effect of 
stray charges: as discussed above, they result in a position dependent
quasiparticle potential energy which has very little effect because  
these quasiparticles had no kinetic energy and were localized anyway. 
A direct effect of stray charges on the topologically protected subspace can be
also physically described as a effect of the electrostatic potential on the states 
with even and odd charges at the inner boundary;
 since the absolute value of the charge
fluctuates strongly this effect is exponentially weak.

Finally, we remark that the properties of the excitations and topological order parameter 
exhibited by this system are in many respects similar to the properties of the ring exchange 
and frustrated magnets models discussed recently in 
\cite{Wen1990,Wen1991,Read1989,Read1991,Kivelson1989,Senthil2001a,Senthil2001b,Balents2001,Fisher2002,Misguich2002,Motrunich2002,Fendley2002,Ioselevich2002}. 

\section{Conclusion} 

We have shown a Josephson junction array of a special type (shown in Fig.~1) 
has a degenerate ground state described by a topological order parameter. The manifold of 
these states is protected in the sense that local perturbations have exponentially weak effect
on their relative phases and transition amplitudes. The main building block of the array is
the rhombus which has two (almost) degenerate states, in the array discussed here these rhombi
are assembled into hexagons but we expect that lattices in which these rhombi form other
structures would have similar properties. However, the dynamics of these arrays is 
described by quartic (or higher) order spin exchange terms which have larger barrier in
a quasiclassical regime implying that their parameter $r$ is much smaller than in for 
the array considered here. This  makes them more difficult to built in the interesting regime.   

We are grateful to G. Blatter, D. Ivanov, S. Korshunov, A. Larkin, A. Millis, B. Pannetier 
and E. Serret for the discussions and useful comments and to LPTMS, Orsay and to LSI, 
Ecole Polytechnique for their hospitality which allowed this work to be completed. 
LI is thankful to K. Le Hur for inspiring discussions. 
MF was supported by SCOPES program of Switzerland, Dutch Organization for Fundamental 
research (NWO), RFBR grant 01-02-17759, 
the program ``Quantum Macrophysics'' of RAS and Russian Ministry of Science.

\end{document}